\documentclass[slac_one]{revtex4}
\usepackage{graphicx}
\usepackage{fancyhdr}
\usepackage{subfigure}
\input babarsym

\begin{document}
\newcommand{\BABARPubYear}    {08}
\newcommand{\BABARPubNumber}  {103}
\newcommand{\SLACPubNumber} {13427}

\begin{flushleft}
BABAR-PROC-\BABARPubYear/\BABARPubNumber \\
SLAC-PUB-\SLACPubNumber
\end{flushleft}

\title{Recent Searches for Exotic Physics at the BaBar/PEP-II B-factory} 

%

\author{Stephen J. Sekula\\(Representing the \babar\ Collaboration)}
\affiliation{The Ohio State University, Columbus, OH 43210, USA}

\begin{abstract}
I present three recent results from searches for exotic physics at the
\babar/PEP-II B-factory. These results span many of the
samples produced at the B-factory, including \B\ mesons,
$\tau$ leptons, and \Y3S\ mesons. We have searched for CPT-violation
in \Bz\ mixing and
find no significant deviation from the no-violation
hypothesis. We have also searched for lepton-flavor-violating decays
of the $\tau$ using $\taum \to \omega \ell^{-}$ and $\taum \to
\ell^{-} \ell^{+} \ell^{-}$ and their charge conjugates. 
We find no evidence for these processes
and set upper limits on their branching fractions. Finally, we have
searched for a low-mass Higgs boson in the decay $\Y3S \to \gamma
A^{0}$, where the Higgs decays invisibly. We find no evidence for such
a decay and set upper limits across a range of possible Higgs masses.
\end{abstract}

\maketitle

\thispagestyle{fancy}


\section{INTRODUCTION} 

The Standard Model of particle physics has been an extremely
successful description of nature. The discovery of neutrino mass, dark
matter, the large matter/anti-matter asymmetry in the universe, and
other natural phenomena which cannot be accommodated by 
the Standard Model require a more comprehensive theory of
nature. 

CPT-conservation is an important principle on which an
effective field theory, such as the Standard Model, is
built. Observation of CPT violation would require
significant changes in our understanding of nature. The
large matter/anti-matter asymmetry in our universe requires sources of
symmetry violation which are not yet understood, such as the violation
of lepton number conservation. Finally, attempts to extend the
Standard Model via supersymmetry (e.g. the Minimal Supersymmetric
Standard Model, or MSSM) succeed in improving the behavior of the
model above the weak scale but introduce new parameters whose values
are not determined by the next natural scale, the Planck scale. 

I report on recent searches at the \babar/PEP-II B-factory for new
physics, including a search for CPT-violation, searches for lepton
flavor violation, and a search for a low-mass Higgs boson which is
produced in the decay $\Y3S \to \gamma A^{0}$. A detailed description of the \babar\ detector can be found elsewhere
\citep{babarnim}.

\section{A SEARCH FOR CPT VIOLATION IN \boldmath{$\B$}-MIXING}

CPT conservation requires that the rate of mixing from $\Bz \to \Bz$
and the rate of mixing from $\Bzb \to \Bzb$  be equal. A violation
of CPT invariance in mixing is detectable as a non-zero value for the
quantity:
\begin{equation}
\mathcal{A}_{CPT} = \frac{Prob(\Bzb\to\Bzb) - Prob(\Bz\to\Bz)}{Prob(\Bzb\to\Bzb) + Prob(\Bz\to\Bz)}
\end{equation}
over the course of a sidereal day.

Our measurement of the CPT asymmetry in \Bz\ mixing is detailed
elsewhere \citep{babarcpt}. I briefly describe our approach and results.
We measure the CPT asymmetry using dilepton events in $232\times
10^{6}$ \BB\ pairs. 
The charge of the lepton indicates the
flavor of the \B\ at the time of decay, and the difference in the
decay positions ($\Delta z$)  of the two \B\ mesons is related to the
time between their respective decays, $\Delta z = \beta\gamma c \Delta t$. 
 
In a generic extension of the Standard Model \citep{Kostelecky_1} a term
$z$ modifies the \Bz\ mass eigenstates \citep{Kostelecky_2}, 
where $z = \beta^{\mu} \Delta a_{\mu} / (\Delta m - i
\Delta\Gamma/2)$, $\beta^{\mu}$ is the boost direction of the
\Y4S\ system, $\Delta a_{\mu}$ is the four-vector of the CPT-violating
effect, and $\Delta m$ and $\Delta \Gamma$ are the mass and
width differences of the heavy and light \Bz\ mass eigenstates.

We search for CPT-violation by measuring the relative numbers of
opposite-sign dilepton events as a function of the sidereal day. The
\B-factory is fixed to the earth, which rotates on its axis and around
the sun. We fix the celestial coordinate system on the center of the
sun. The z-axis ($\hat{Z}$) of the celestial
coordinate system is parallel to the earth's rotation axis, and the
z-axis ($\hat{z}$) of the rotating frame is opposite the
\Y4S\ boost vector. The axis $\hat{z}$ precesses around $\hat{Z}$ with
a period of one sidereal day. 

\begin{figure}
\subfigure[\label{siderealtime} CPT asymmetry vs. sidereal time for
  opposite-sign dilepton events. The curve is a projection of the
  two-dimensional maximum likelihood fit.]{
  \includegraphics[height=5cm]{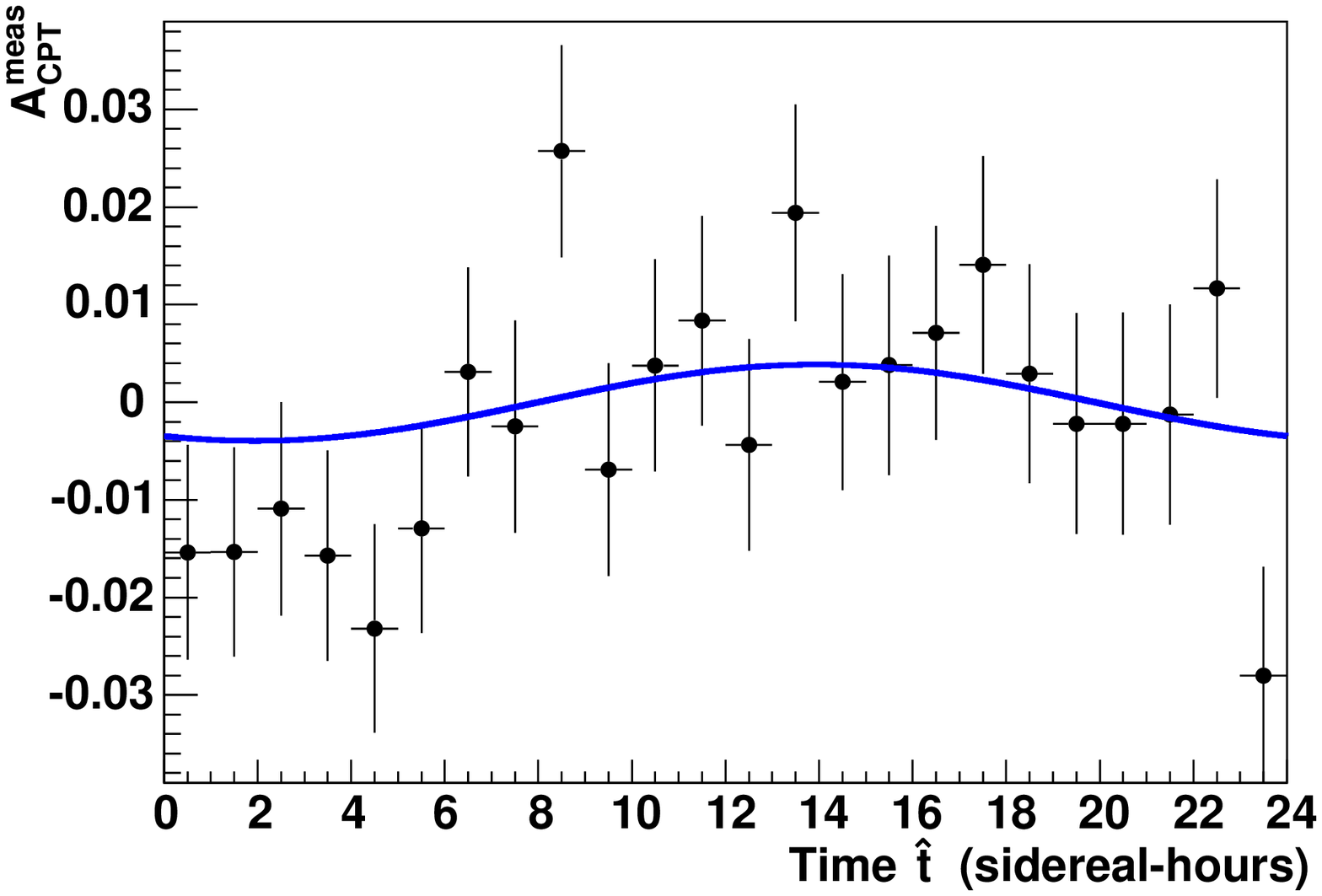}%
}%
\subfigure[\label{periodigram} Periodogram for opposite-sign dilepton
  events. The solar day and sidereal day frequencies, indicated by the
left and right triangles (inset), are well-resolved in our data.]{
  \includegraphics[height=5cm]{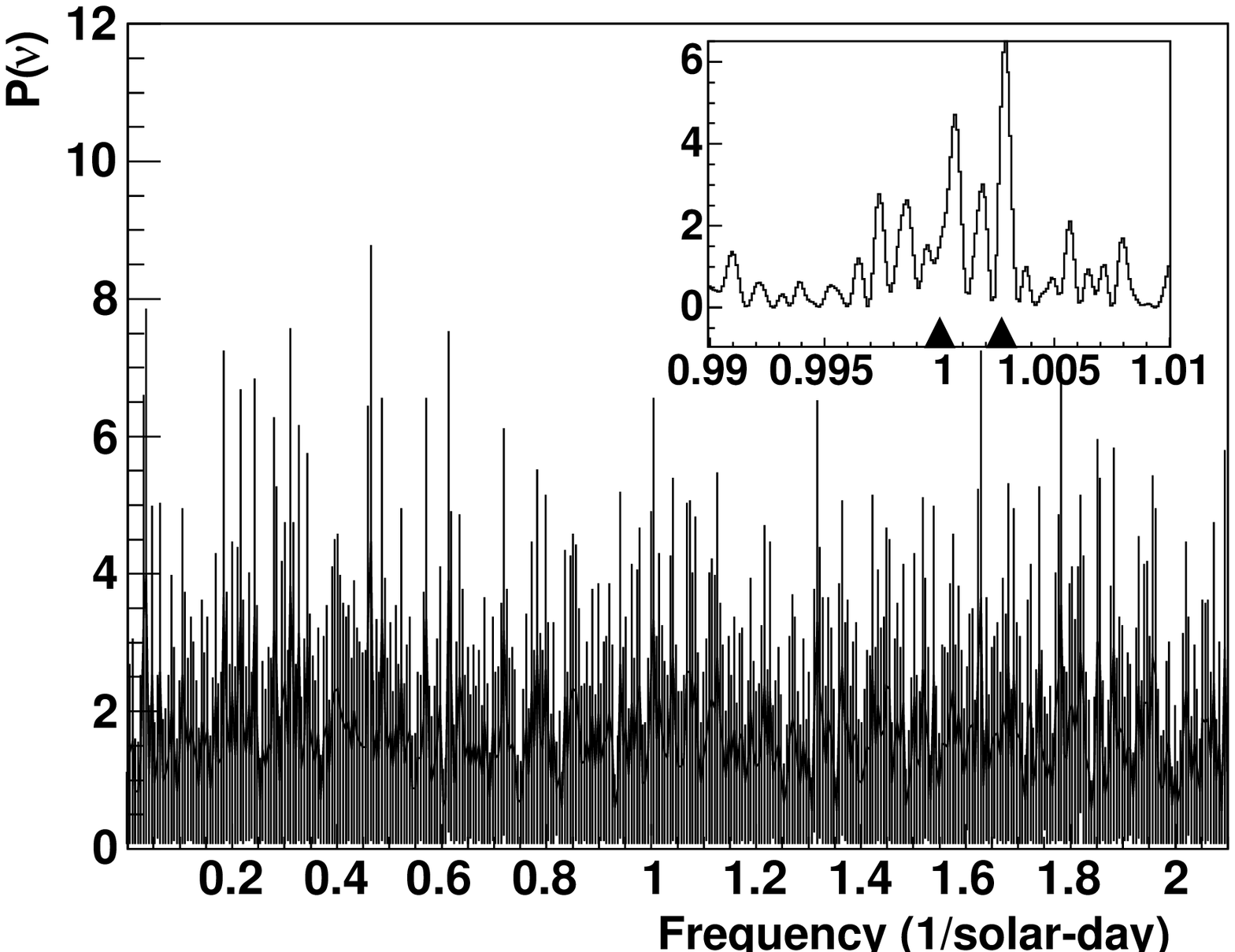}%
}%
\caption{Results from the two methods of analyzing the \Bz\ mixing data}
\end{figure}

The dilepton events are fitted in two dimensions - $\Delta t$ and
sidereal time - using a maximum likelihood function. The projection of
the fit and data in only sidereal time is shown in
Fig. \ref{siderealtime}. We
find that the data are consistent with the no-CPT-violation hypothesis
at the level of $2.8\sigma$.

We perform an alternative measurement of the data by studying the spectral power of
periodic variations in $z$ over a wide frequency band using the
periodogram method~\cite{pgram} developed to study variable stars. The
spectral power at a test frequency $\nu$ is $P(\nu) \equiv \frac{1}{N\sigma^2_w}\Bigl|\sum_{j=1}^{N}w_j e^{2i\pi\nu T_j}\Bigr|^2$,
where the data, containing $N$ measurements of the weight, $w_j$, made
at times $T_j$, with variance $\sigma^2_w$. 
Here, $T_j$ is the time elapsed since the Unix epoch for 
opposite-sign dilepton event $j$, and the weights 
$w_j = \Delta m \Delta t_j - \sin(\Delta m \Delta t_j)$ are suited to
the study of periodic variations in $z$ according as described by
$\mathcal{A}_{CPT}$. 

The frequency spectrum is shown in Fig. \ref{periodigram}. The largest
spectral power occurs at nearly half a sidereal day, and assuming no
signal we find that the probability of obtaining a larger spectral
power is 76\%. We find that for events with a frequency of one
sidereal day, there are 78 other frequencies (out of 20994 test
frequencies) which exceed its spectral power.  

The dominant source of uncertainty is due to statistical uncertainty. Therefore, with the remaining half of the
\babar\ dataset we can expect to improve this measurement by reducing
the dominant statistical uncertainties.

\section{SEARCHES FOR LEPTON-FLAVOR-VIOLATING $\boldmath{\tau}$ DECAYS}

We use a sample of $\sim 700 \times 10^{6}$ $\tau$ leptons
to search for the lepton-flavor-violating (LFV) decays $\taum \to \omega \ell^{-}$ and $\taum \to
\ell^{-} \ell^{+} \ell^{-}$ (charge conjugation is implied throughout). 
Observation of either is an unambiguous
sign of physics beyond the Standard Model.

These analyses are described in detail elsewhere \cite{taulfv}. I
briefly describe our approach and results. We reconstruct the signal
events by separating the event into two hemispheres based on the
thrust axis. One hemisphere is required to contain only one charged
particle identified as either an electron or muon, while the other is
required to contain three charged particles. We require that the three
particles pass muon or electron identification for the $\taum \to
\ell^{-} \ell^{+} \ell^{-}$ final state, while for the $\taum \to
\omega \ell^{-}$ final state one track must be identified as either an
electron or muon, and the other two must be paired with a $\piz$
candidate to form an $\omega$ candidate.

We find, using Monte Carlo simulations, that the background for each
final state is expected to be about one event and that the efficiency
for reconstructing the signal varies with final state between
$2-10\%$. We find no evidence for these decays in our data and set
upper limits on the branching fractions for these decays. Depending on
which combination of lepton identification is used, the limits on the
branching fraction for $\taum \to \ell^{-} \ell^{+} \ell^{-}$ range
between $(4-8) \times 10^{-8}$. For both lepton final states of $\taum \to
\omega \ell^{-}$, the limits are $1 \times 10^{-7}$. All limits are at
the 90\% confidence level. These results are competitive with the best
existing limits on these decays.

\section{A SEARCH FOR A LOW-MASS, INVISIBLY DECAYING HIGGS BOSON}

We use a sample of $122 \times 10^{6}$ \Y3S\ mesons to search for the
production of a low-mass CP-odd Higgs boson in the decay $\Y3S \to
\gamma A^0$. Such a low-mass Higgs boson is possible if the MSSM is 
extended with an additional gauge singlet Higgs chiral
superfield \citep{nmssm}. The addition of such a singlet yields two additional
Higgs bosons, including a CP-odd Higgs which can have a mass below the
\bbbar\ threshold, and allows the Higgs vacuum expectation value in the
MSSM to be dynamically generated, instead of set by hand to the weak
scale.

While the dominant decay of such a low-mass Higgs boson could be to
$\tau$ leptons, muons, etc. if there is also additionally a low-mass
dark matter component then the dominant Higgs decay may be to a
completely undetectable final state. We therefore search for a single
monochromatic photon recoiling against an invisibly decaying Higgs boson. 

This analysis is described in detail elsewhere \citep{higgs}. I briefly
describe the method and results of this search.
We select events by requiring they contain a single, well-
reconstructed photon within the barrel of the electromagnetic
calorimeter. We require that there are no charged particles in the
event, and we constrain the additional neutral particles to have very
little total energy. The second-highest energy neutral particle cannot
be back-to-back with the signal photon; this is done to eliminate part
of the significant background from $\ep\en\to\gamma\gamma$. There is a
significant contribution from this background where the second photon
escapes detection in the calorimeter but leaves hits in the
instrumented flux return. We require that there are no hits in that
system correlated with the signal photon.

We search for a signal in the data by performing a maximum likelihood
fit to the square of the missing mass, 
\begin{equation}
m^2_X = M^2_{\Y3S} - 2 E_{\gamma} M_{\Y3S}.
\end{equation}
We fit the data using several models for backgrounds (a peaking
component from $\ep\en\to\gamma\gamma$ at $m^2_X = 0 \gev^2$, with a
long tail extending to large values of $m^2_X$, and non-peaking
components from other sources) and a model for the signal. The signal
model represents the resolution function of a photon reconstructed in
the calorimeter, and its parameters vary with energy as the
calorimeter response varies. We scan the signal model across $m^2_X$
in steps of 100\mev\ in $m_X$ for $m_X^2 < 40\gev^2$ and in steps of
25\mev\ for  $m_X^2 > 30\gev^2$. Due to the different triggers
required to select events in these two regions, there is a slight
overlap which is handled in the final combination of results.

\begin{figure}
\subfigure{
  \includegraphics[height=5cm]{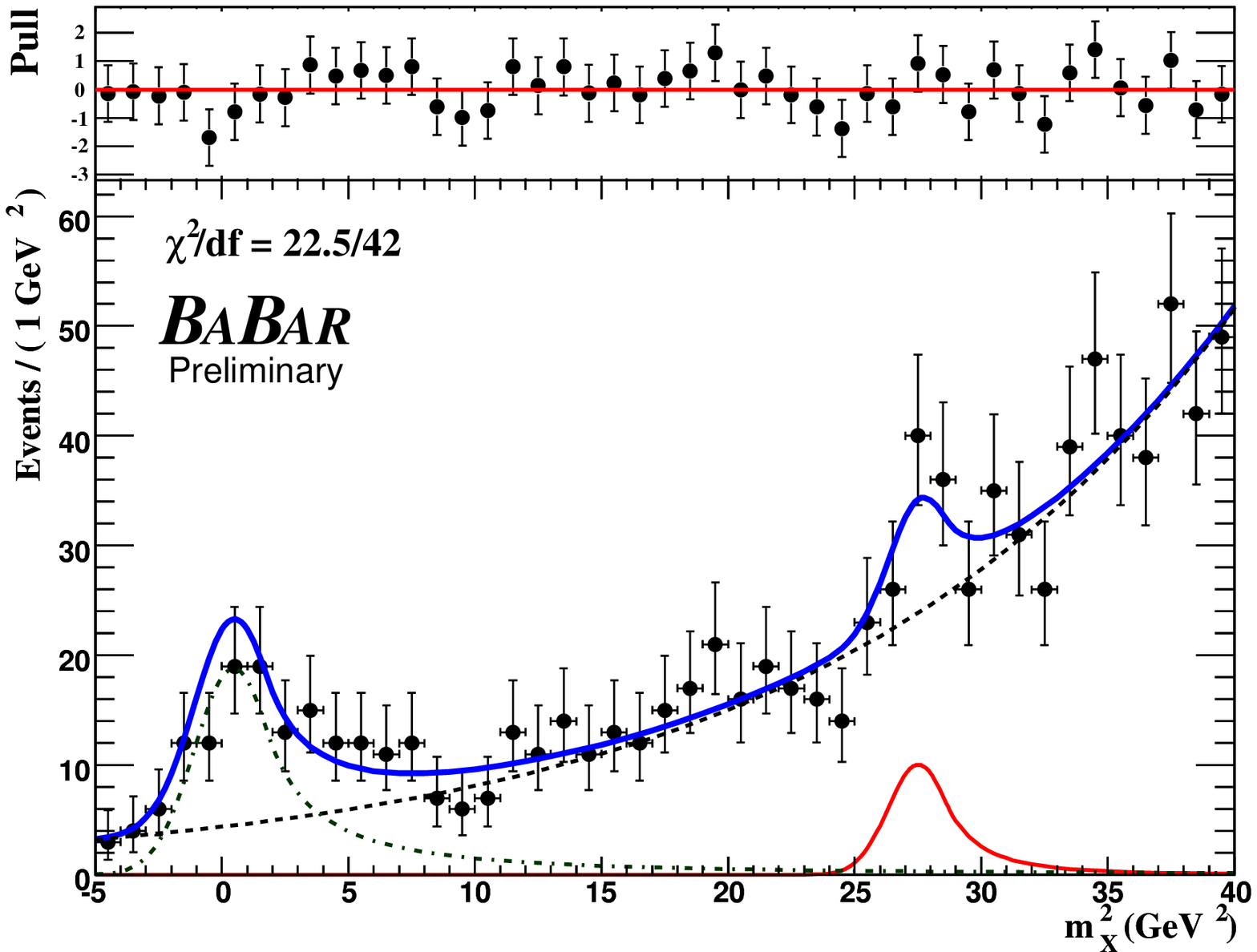}
}%
\subfigure{
  \includegraphics[height=5cm]{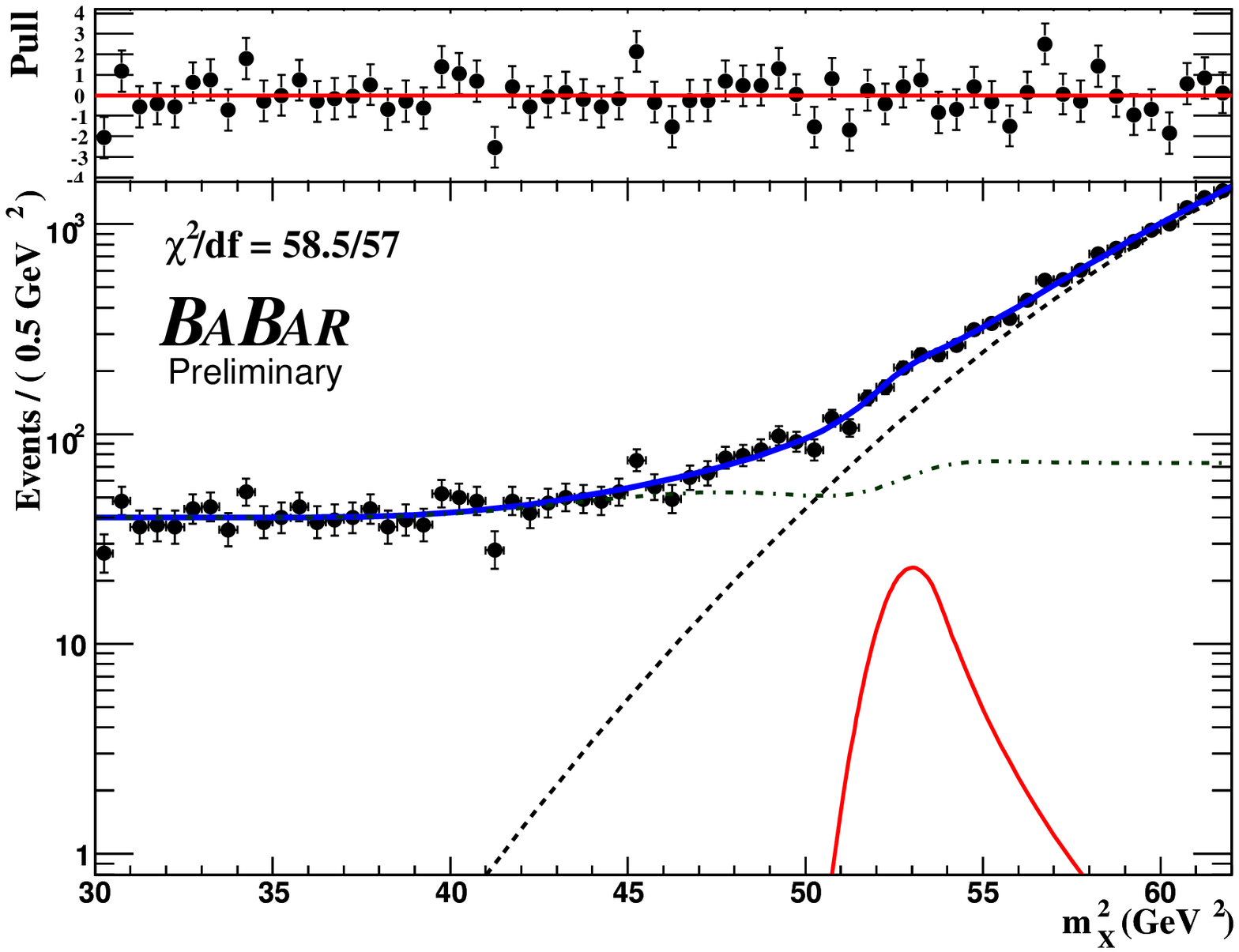}
}%
\caption{\label{higgsfits}Two example fits from the Higgs scan, one
  from the low-missing-mass data (left) and one from the
  high-missing-mass data (right), each representing the most
  significant yields from each region.}
\end{figure}

Two example fits, one from the low-missing-mass region and one from
the high-missing-mass region, are shown in Fig. \ref{higgsfits}. These
examples are chosen because they represent the most significant yields
in each missing-mass region - $(37 \pm 15)$ events in the
low-missing-mass region and $(119 \pm 71)$ events in the
high-missing-mass region. Neither of these is a significant enough
deviation from the null-signal hypothesis, with the larger of the two
statistical significances being $2.6\sigma$.

\begin{figure}
\includegraphics[width=0.6\linewidth]{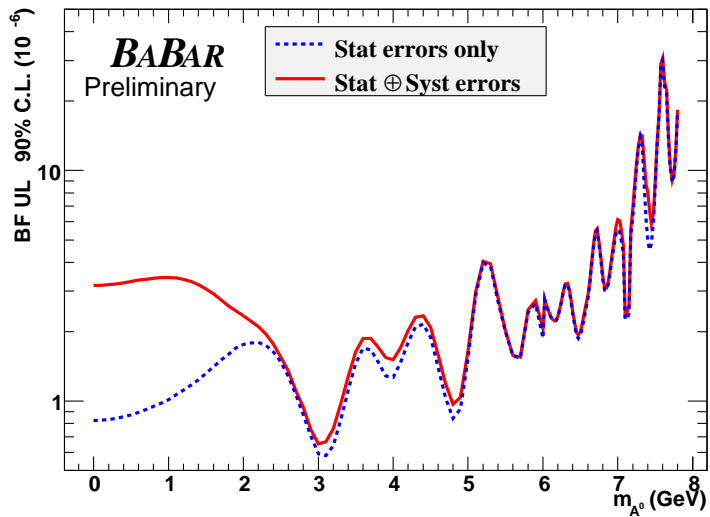}%
\caption{\label{higgslimits}The upper limit on the branching fraction
  for $\Y3S \to \gamma A^{0}$ as a function of Higgs mass. }
\end{figure}

We interpret the results of these fits as upper limits on the
branching fraction for $\Y3S \to \gamma A^{0}$ as a function of the
Higgs mass. The results are shown in Fig. \ref{higgslimits}. The
statistical uncertainty dominates across the mass spectrum, except in
the region where the $\ep\en\to\gamma\gamma$ background peaks. For
small Higgs masses, the systematic uncertainty on the shape and yield
of this background dominates. These are
the most significant constraints on an invisibly decaying low-mass
Higgs boson for a Higgs up to $m_{A^0} = 7.8\gevcc$.

\end{document}